\documentstyle{article}

\newcommand{\mathbf}{\bf}

\begin{document}

\begin{center}
{\huge\bf On a physical realization of Chern-Simons Theory}
\end{center}

\vspace{1cm}
\begin{center}
{\large\bf
F.GHABOUSSI}\\
\end{center}

\begin{center}
\begin{minipage}{8cm}
Department of Physics, University of Konstanz\\
P.O. Box 5560, D 78434 Konstanz, Germany\\
E-mail: ghabousi@kaluza.physik.uni-konstanz.de
\end{minipage}
\end{center}

\vspace{1cm}

\begin{center}
{\large{\bf Abstract}}

The physical content of Chern-Simons-action is discussed and it is  
shown that this action is proportional to the usual chraged matter  
interaction term in electrodynamics.
\end{center}

\begin{center}
\begin{minipage}{12cm}

\end{minipage}
\end{center}

\newpage

During the last decade we had a rapid success of Chern-Simons  
theory in theoretical physics \cite{math}. Almost in every part of  
physics where the topological and low dimensional effects play  
roles, one finds applications of this theory \cite{all}. This  
applications show the physical relevance of Chern-Simons theory in a  
clear manner.

However, there are still two open queastions about the physical  
realization of this theory, namely about:

1. A physical realization of Chern-Simons theory according to the  
well known and observable physical phenomena \cite{c.q}.

2. The physical relationship between the well known  
Maxwell/Yang-Mills theories and the Chern-Simons theory.

We discuss here the first question and show that there is indeed a  
well known
electrodynamical realization of the U(1) Chern-Simons structure  
according to the Ohm's equations.

A generalization to the U(N) Chern-Simons theories can be also  
considered in view of the fact that the gauge $A_0 = 0$ which is  
needed for quantization of the U(N) version reduces also the U(N)  
form into the U(1) form.

To begin let us mention an early result of us which comes out in  
relation with the theory of IQHE \cite{mein}.

If we consider the Chern-Simons action $S_{C-S} = \int d^3 x  
\epsilon^{\alpha \beta \gamma} A_{\alpha} \partial_{\beta}  
A_{\gamma}$ in the usual Weyl-gauge $A_0 = 0$, then the action  
becomes \cite{witt}

\begin{equation}
S = \int_M \epsilon_{mn} A_m \frac{dA_n}{dt}\;\;\;, m,n = 1,2\;\;\;,
\end{equation}
\label{ein}

where $M:= \Sigma\times {\bf R}$ is the three dimensional  
space-time manifold.

If we insert the Ohm's equations in two dimensions
$j_m = \sigma_H \epsilon_{nm} \displaystyle{\frac{dA_n}{dt}}$ with  
$\epsilon_{mn} = - \epsilon_{nm} = 1$ which are known, as  
phenomenological relation or "material equations", from the quantum  
Hall effect with $j_m = n e v_m$ and $\sigma_H$ as the current  
density and the Hall conductivity respectively, where $n$ is the  
charge carrier density and $e$ is the elementary charge. Then, the  
action (1) becomes proportional to the following one

\begin{equation}
Q \int A_m dx^m \;\;\;,
\end{equation}
\label{zwei}

which is well known from the classical electrodynamics for  
interaction term between potential and charged matter \cite{landau},  
where $ C = 1$ and $Q = \int\int ne$ and the proportionality factor  
is the Hall resistivity.

Thus, it is obvious that one may obtain in the opposite way the  
Chern-Simons electrodynamics from the standard interaction term  
between the charged matter and electromagnetic potential in the  
classical action functional \cite{landau}.

\bigskip
To arrive at the physically general relevant form one should begin  
with the usual interaction term
$Q \int A_{\mu}dx^{\mu}$
between a charged matter and electromagnetic potential in classical  
theories where $\mu ={\{0,...,3}\}$.

Considering a continuous 3-D charge distribution one obteins  
\cite{landau}:

\begin{equation}
Q \int A_{\mu}dx^{\mu} = \int d^4x A_{\mu}j^{\mu} \;\;\;,
\end{equation}
\label{drei}

where we used the definition of current density $j_{\mu} = ne v_{\mu}$.

In view of the continuity equation for $j_{\mu}$ or in view of the  
gauge dependence of electromagnetic potential we have to choose  
gauge fixing condition(s), e. g. $j_0 = 0$ or $A_0 = 0$ to arrive at  
physically relevant interaction term \cite{gauge}. This reduces the  
above mathematically general invatiant term to the following  
physical general invariant interaction term

\begin{equation}
\int dt \int d^3 x A_{\alpha}j^{\alpha} \;\;\;, \;\;\; \alpha ={1, 2, 3}
\end{equation}
\label{vier}

Recall that , if we choose the gauge fixing condition $j_3 = 0$ or  
$A_3 = 0$, then $\alpha ={\{0, 1, 2}\}$ and $dt$ must be replaced by  
$dz$.

Thereafter, we can rewrite it in view of Ohm's equations  
$j_{\alpha} = \sigma_H \epsilon_{\alpha \beta \gamma}  
\partial_{\beta} A_{\gamma}$ in the following invariant form

\begin{equation}
\sigma_H \cdot \int dt \int d^3 x \epsilon^{\alpha \beta \gamma}  
A_{\alpha} \partial_{\beta} A_{\gamma} \;\;\;,
\end{equation}
\label{funf}

where $\sigma_H$ is a locally constant quantity called the Hall  
conductivity.

The above volume integral is the well known Chern-Simons invariant  
in $3$ or $2+1$-D manifold. Thus, we find a very general physical  
realization of the Chern-Simons action functional in terms of the  
very usual interaction term between matter and the electromagnetic  
potential.

Here, the volume integral, i.e. the Chern-Simons-action can be  
considered as the interaction Lagrangian between charged matter and  
the electromagnetic potential.

If we choose for the second gauge fixing condition $A_0 = 0$ and  
use $A_m:= A_m (x_m, t)$, then the relation reduces to the one  
proportional to the relation (1), where the proportionality factor  
is a constant times $\sigma_H$.

Moreover, if we start with a $2-D$ charge distribution in the  
interaction term (3), then we arrive at $Q_{(2-D)} \int A_{\mu}  
dx^{\mu} = \sigma_H \int d^3x \epsilon^{\alpha \beta \gamma}  
A_{\alpha} \partial_{\beta} A_{\gamma}$.

Thus, the question is if the dimension of charge distribution  
should be related with the true degrees of freedom of the  
electromagnetic field or of photon which is the intermediating field  
of electromagnetic interaction, i. e. also responcible for charged  
particle. This question, however should be discussed seperately  
\cite{under}.

\medskip
It is intresting to mention that the usual Weyl gauge fixing $A_0 =  
0$ which reduces the general Chern-Simons action $\int A \wedge dA$  
in $2+1-D$ to its {\it quantizable} form:

\begin{equation}
\int dt \int d^2x \epsilon_{mn} A_{m} \partial_t A_{n} \;\;\;,
\end{equation}
\label{sechs}

according to the $[ A_m (X, t) , A_n (Y, t)] = i \hbar  
\epsilon_{mn} \delta^2 (X - Y); X,Y \in \Sigma $ postulate  
\cite{witt}, reduces also the number of degrees of freedom of  
electromagnetic field which was originally three to its physically  
true degrees of freedom, i. e. to two.

This is a hint about the physical relevance of the Chern-Simons  
action in the sence that the quantizable Chern-Simons action, i. e.  
action (6) depends only on the physically relevant degrees of  
freedom of the electromagnetic field, whereas in case of Maxwell's  
action there is no such relation between the quantization conditions  
on this action and the true degrees of freedom of electromagnetic  
field. The reason is of course the different structures of action in  
the two cases and that the U(1) field strength is gauge invariant.

Moreover, the constraints of Chern-Simons action $\epsilon_{mn}  
\partial_m A_n = 0$ forces the electromagnetic potential to be a  
pure gauge potential. On the other hand, it is known from the  
geometric quantization that the quantization is caused by the  
existence of a flat connection on a line bundle over the phase space  
\cite{wood}. In the case under consideration, it is the pure  
electromagnetic gauge potential which should play the role of the  
mentioned flat connection in geometric quantization, where the  
related U(1) bundle of gauge transformation, as a pricipal bundle,  
should replace the mentioned equivalent line bundle.

\medskip
This conception is also in accordance with the Borel-Weil  
construction of the Hilbert spaces as the space of holomorphic  
sections of a complex line bundle $L_{\lambda}$ over the $G/T$ where  
$G$, $T$ and $\lambda$ are a compact finite dimensional group, its  
maximal Abelian subgroup and the highest weigth of an irreducible  
representation of $G$ \cite{qft}.

Thus, the quantized Chern-Simons action in its Schroedinger  
representation $\phi = e^{i \sigma_H S_{C_S}} = e^{iQ \int_C A_m  
dx^m}$ represents the phase change of a charged system moving in an  
electromagnetic potential field:

\begin{equation}
\Psi_c ^A (x_m, t) = e^{iQ \int_C A_m dx^m} \Psi (x_m, t)
\end{equation}
\label{funf}

which is a path dependent gauge transformation of $\Psi (x_m, t)$.

The path independent phase change or gauge transformation is then  
given by a closed path contribution which is equivalent to the  
Bohm-Aharonov effect for $B \neq 0$, i. e. in a multiply connected  
region. Followingly, this could be considered as an observable  
physical realization of Chern-Simons-theory, if one recalls the  
Ohm's equation for charge carrieres. A related quantum mechanical  
realization of Chern-Simons theory according to the above relation  
with $\int A$ should be seen in the flux quantization in view of  
$\oint A = \displaystyle{\frac{h}{e}}$.

An other application of the mentioned relation between  
Chern-Simons-action and the usual interaction term is that one can  
show the equivalence between the Kubo-Thouless approach to the IQHE  
which is based on the linear response theory \cite{all} and the  
Chern-Simons-approach to the same \cite{mein}. The Kubo's linear  
response theory use as perturbation term the $ H_t = A_{\mu}  
j^{\mu}$. Thus, as it is shown already the perturbative action will  
be in the $2+1$ dimensional case, in view of Ohm's equations, equal  
to the Chern-Simons-action.
Recall that in the linear response approach the conducticity is  
identified by comparison with other phenomenologically known  
formula, thus it is identified here phenomenologically. This is  
equivalent to the use of Ohm's equations as phenomenological  
equations, in the above discussed relation between two action terms.

This means that in case of theory of conductivity the Chern-Simons  
interaction term and linear response interaction term are both  
equivalent to the usual interaction term $A_m dx^m = A$.

Moreover, the related prove of antisymmetry of conductivity in IQHE  
case according to the linear response approach is a consequence of  
antisymmetry of $F_{mn}$ in the integral relation $\oint A =  
\int\int F$, with $F_{mn} := [ D_m, D_n ]$ and $D_m = \partial_m -  
A_m$. The reason lies in the flux quantization $\oint A = \int\int F  
= \displaystyle{\frac{h}{e}}$ which should be responsible also for  
the quantization in IQHE case \cite{under}.

\medskip
In conclusion let us mention that also the mentioned second  
question should be answered according to the above discussed point  
of view, if one recalls the relation between the Chern-Simons-action  
and the Adler-Bell-Jakiw anomaly term:

\begin{equation}
Tr \int dA \wedge dA = Tr \; (d \int A \wedge dA) \;\;\;
\end{equation}
\label{sechs}

which arises in quantization of Yang-Mills action \cite{under}.

\bigskip
Footnotes and references


\begin{thebibliography}{100}
\bibitem{all}
Beyond the application of Chern-Simons Field theory in the quantum  
Hall Effect (IQHE and FQHE), one has also applications in  
superconductivity, Chaos theory, etc. (see Ref. \cite{mein} and  
references therein for related applications)
\bibitem{c.q}
It could be a realization according to the classical or quantum  
physics. Nevertheless, it is importent to mention that the classical  
interpretation of Chern-Simons theory according to the  
electromagnetic interaction term for charged matter which is  
discussed here, have its quantum mechanical realizetion, e.g.  
according to the Bohm-Aharonov effect.
\bibitem{gauge}
In view of the gauge dependence of the electromagnetic potential  
$A^{\prime}_{\mu} = A_{\mu} + \partial_{\mu}\Lambda$ one has to  
choose two gauge fixing conditions, e.g. the radiation or Coulomb  
gauge, to retain the two true degrees of freedom for the  
electromagnetic potential or photon.
\end{thebibliography}
\end{document}